\documentclass[fleqn,twoside]{article}
\usepackage{espcrc2}

\usepackage{graphicx}
\usepackage[figuresright]{rotating}

\newcommand{\AmS}{{\protect\the\textfont2
  A\kern-.1667em\lower.5ex\hbox{M}\kern-.125emS}}

\title{Mass Hierarchy via M\"ossbauer and Reactor Neutrinos}

\author{\bf{Stephen~J.~Parke,\address{Theoretical Physics Dept.,
Fermi National Accelerator Laboratory, 
Batavia, IL 60510, USA }\thanks{Presenter. email:parke@fnal.gov}
H.~Minakata,\address{Department of Physics, Tokyo Metropolitan University, 
Hachioji, Tokyo 192-0397, Japan}
H.~Nunokawa,\address{Dept. de F\'{\i}sica, Pontif{\'\i}cia U. Cat{\'o}lica 
do Rio de Janeiro, C. P. 38071, 22452-970, Rio de Janeiro, Brazil }
R.~Zukanovich Funchal\address{Instituto de F\'{\i}sica, Universidade de S\~ao Paulo, 
 C.\ P.\ 66.318, 05315-970 S\~ao Paulo, Brazil}}}
       
\begin{document}

\begin{abstract}
We show how one could determine the neutrino mass hierarchy with M\"ossbauer neutrinos and also revisit the question of whether the hierarchy can be determined with reactor neutrinos.
\vspace{1pc}
\end{abstract}

\maketitle

\section{M\"ossbauer Neutrinos}
The term M\"ossbauer Neutrinos refers to a source-detector combination where there is no recoil
in the production and absorption of the neutrino so that the neutrinos have resonant energy
thereby enhancing the capture cross section by approximately 10 orders of magnitude!
\begin{eqnarray}
{\rm Source:} \quad ^3H \rightarrow (^3He+e^-)_B +\bar{\nu}_e \nonumber \\
{\rm Detector:} \quad \bar{\nu}_e+ (^3He+e^-)_B \rightarrow ^3H \nonumber
\end{eqnarray}
Such a source-detector combination would be wonderful for neutrino physics if it could be practically realized, see \cite{Raghavan:2006xf}. Here we address the question of whether such a system could be used to determine the neutrino mass hierarchy.

\section{$\nu_e$ Disappearance Probability}

\begin{figure}[bhtp]
\includegraphics[width=0.45\textwidth]{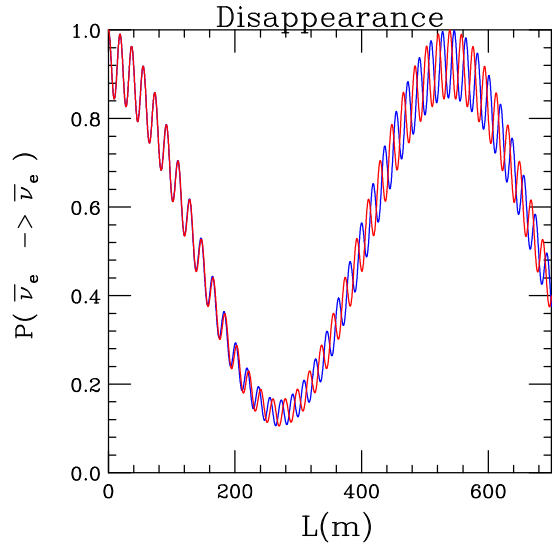}
\vspace*{-1cm}
\caption{The antineutrino survival probability
  $P(\bar{\nu}_e \rightarrow \bar{\nu}_e)$ is plotted as a function of
  $L$ for the 18.6 keV M\"ossbauer Neutrinos for both the normal (blue) and inverted (red) hierarchies.
  The advancement/retardation in the phase of the atmospheric oscillation is clearly visible beyond the first solar oscillation minimum.  Here the quantity $\Delta m^2_{ee}  \equiv c^2_{12} \vert \Delta m^2_{31} \vert + s^2_{12} \vert \Delta m^2_{32} \vert $ is the same for both hierarchies. 
  }
\label{fig:disapp}
\end{figure}

The vacuum $\nu_e$ survival probability using 
$\Delta_{ij} \equiv \Delta m^2_{ij} L/4E$ ($\Delta m^2_{ij}\equiv m^2_i-m^2_j$)
as shorthand for the kinematical phase,
can be written without any approximation as
\begin{eqnarray}
P(\nu_e \rightarrow \nu_e)  = 1- 
\cos^4 \theta_{13} \sin^2 2\theta_{12} \sin^2 \Delta_{21} 
\nonumber  \\ 
-\sin^2 2 \theta_{13}  \left[
\cos^2\theta_{12} \sin^2 \Delta_{31} + \sin^2\theta_{12} \sin^2 \Delta_{32}  
\right]. \nonumber 
\label{Pee}
\end{eqnarray}
\begin{figure}[tb]
\includegraphics[width=0.45\textwidth]{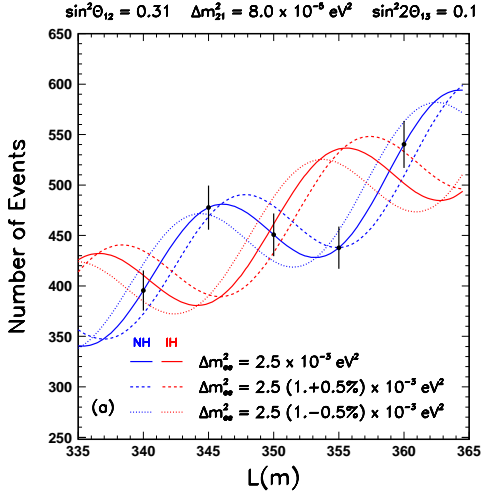}
\caption{Plotted are the expected number of events to be collected by detectors 
placed at the distances 340, 345, 350, 355 and 360 m from the source
for $\Delta m^2_{{\rm ee}}=2.5 \times 10^{-3}$ eV$^2$ 
and $\sin^2 2\theta_{13} = 0.1$ for the normal hierarchy, 
indicated by solid circles with error bars. 
}
\label{fig:disapp2}
\end{figure}

The first term gives the contribution from the solar $\Delta m^2$ whereas the last two terms 
are the contributions from the two atmospheric $\Delta m^2$s.  This atmospheric contribution consists
of two waves with slightly different frequencies which leads to
\begin{itemize}
\item a modulation of the amplitude of the atmospheric oscillations by the factor $\sqrt{1-\sin^2 2\theta_{12} \sin^2 \Delta_{21}}$
\item and an advancement  (retardation) of the phase of the atmospheric oscillation
by $2\pi \sin^2 \theta_{12}$ for every solar oscillation ($\Delta_{21} \rightarrow \Delta_{21}+\pi$) assuming the normal (inverted) hierarchy.
\end{itemize}

The easiest way to see these two effects is to combine the two atmospheric oscillation terms into
a single term as follows:
\begin{eqnarray}
2\left[\cos^2\theta_{12} \sin^2 \Delta_{31} + \sin^2\theta_{12} \sin^2 \Delta_{32} \right] 
 =  \\
\left[ 1-  \sqrt{1- \sin^2 2 \theta_{12} \sin^2 \Delta_{21}}  
~\cos (2 \Delta_{\rm ee} \pm \phi_\odot) \right]  \nonumber
\label{eqn:Pdis}
\end{eqnarray}
where the argument of the new cosine term, $(2 \Delta_{\rm ee} \pm \phi_\odot)$, has been separated into a linear term in L/E, $2\Delta_{ee} \equiv \Delta m^2_{ee} L/2E$, and the remainder, $\pm \phi_\odot$, whose derivative wrt L/E vanish at L/E=0.

The $\Delta m^2$ associated with the linear term, $2\Delta_{\rm ee}$, is given by
\begin{eqnarray}
\Delta m^2_{ee} & \equiv & c^2_{12} \vert \Delta m^2_{31} \vert + s^2_{12} \vert \Delta m^2_{32} \vert \\
& = & \vert m^2_3-(c^2_{12} m^2_1+s^2_{12} m^2_2)\vert  \nonumber
\end{eqnarray}
and is the electron flavor weighted average of   $\vert \Delta m^2_{31} \vert$ and $\vert \Delta m^2_{32} \vert$. $ \Delta m^2_{ee}$  is the atmospheric $\Delta m^2$ measured by a electron neutrino disappearance experiment in the first few oscillations from the source. 

The $\pm \phi_\odot$ contains everything else, with the positive (negative) sign for the normal
(inverted) hierarchies, and only depends on the solar $\Delta m^2$ as follows
\begin{eqnarray}
\phi_\odot \equiv  \arctan(\cos 2 \theta_{12} \tan\Delta_{21})-\Delta_{21} \cos 2\theta_{12}.
\end{eqnarray}
$\phi_\odot$ is a monotonically increasing function of $\Delta_{21}$ and changes by
$2\pi \sin^2 \theta_{12}$ for every $\pi$ that $\Delta_{21}$ increases, i.e. $\phi_\odot(\Delta_{21}+\pi)=
\phi_\odot(\Delta_{21})+2\pi \sin^2 \theta_{12}$.

\begin{figure}[tbh]
\vglue -0.2cm
\includegraphics[width=0.45\textwidth]{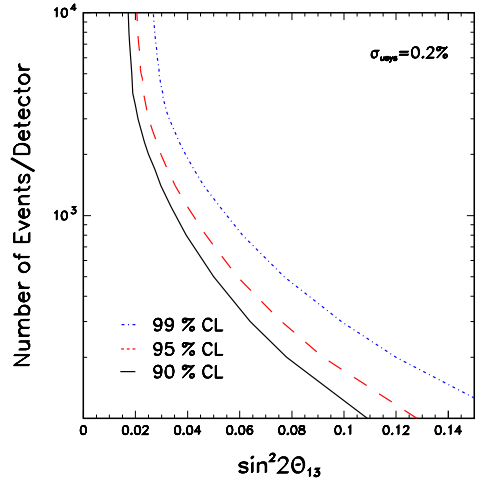}
\vglue -0.6cm
\caption{The region of sensitivity to resolving the mass hierarchy in 
$\sin^2 2\theta_{13} -$event number (per detector) space. 
The black solid, the red dashed, and the blue dotted curves 
denote the region boundary at 90\%, 95\%, and 99\% CL, respectively. 
The uncorrelated systematic uncertainties are assumed to be of 0.2\%. 
}
\label{fig:sensitivity}
\end{figure}


It is this advancement or retardation of the phase of the atmospheric oscillation which can exploited to 
determine the mass hierarchy using only the $\nu_e$  vacuum disappearance channel.
Thus, the strategy is as follows: to make a precise determination of the atmospheric $\Delta m^2$
near the first few atmospheric oscillation and then to go out beyond the first solar minimum 
and determine the phase of the atmospheric oscillation at the point where the atmospheric oscillations from normal and inverted hierarchy scenarios are close to 180 degrees out of phase,  see Figures \ref{fig:disapp} and \ref{fig:disapp2}.
Since this point is 20 atmospheric oscillation from the source the precise determination of the $\Delta m^2$ (better than 1\%) near the source is required so that one can determine whether the measured phase is associated with the normal or inverted hierarchy.  Also the L/E of the events is required to be determined by better than 1\% or otherwise the atmospheric oscillations are averaged out. For an ultra-monochromatic neutrino source like M\"ossbauer Neutrinos this is not a very stringent requirement 
provided that the source and detector are relatively compact.
Figure \ref{fig:sensitivity} gives the sensitivity to determining the mass hierarchy assuming a certain number of events per detector as discussed in \cite{Minakata:2007tn}.

\begin{figure}[tbh]
\vglue -0.2cm
\includegraphics[width=0.45\textwidth]{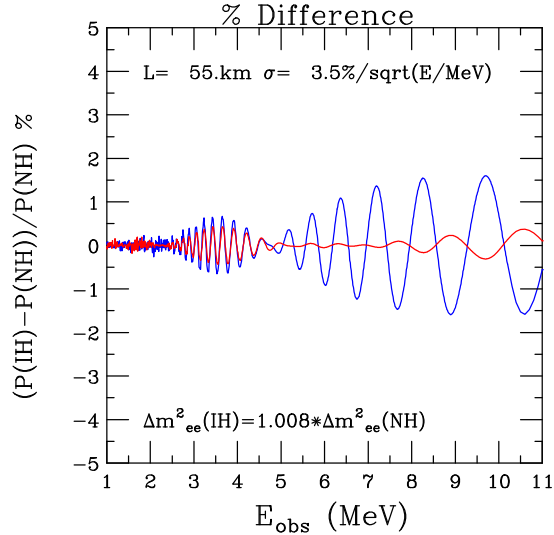}
\vglue -0.6cm
\caption{The percentage difference between the inverted hierarchy and the normal hierarchy. The blue curve is assuming $E_{obs} = E_{true}$ and maximum difference is less than 2\%. Whereas for the red curve we have assumed that $E_{obs} = 1.015E_{true}-0.07$~MeV for the IH, so as to represent a relative calibration uncertainty in the neutrino energy. Here the maximum percentage difference is less than 0.5\%.}
\label{fig:hano}
\end{figure}
We now turn to the question of whether reactor neutrinos can be used to determine the neutrino mass
hierarchy using the difference in the disappearance probability for the normal and inverted hierarchies.
This issue has been discussed in some detail in a recent paper with respect to the Hanohano experiment, see \cite{Learned:2006wy}. In Fig. \ref{fig:hano} we have plotted the percentage difference in the disappearance probability assuming 
\begin{eqnarray}
\Delta m^2_{ee}(IH) =1.008\times \Delta m^2_{ee} (NH)
\end{eqnarray}
 with this choice the difference between the two hierarchies is minimized  in the energy window 2-8 MeV accessible with reactors.
 If we know the energy of the neutrinos exactly, $E_{obs} = E_{true}$, then the difference between
 the two hierarchies is approximately 1\%.
 
 However, if the measured neutrino energy differs from the true energy by a small amount, say
 \begin{eqnarray}
 E_{obs} = 1.015E_{true}-0.07 {\rm~MeV}, 
 \end{eqnarray}
 then the difference between the inverted hierarchy oscillation probability using $E_{obs}$ and the normal hierarchy with $E_{true}$ can be considerable smaller than 1\%. 
 Thus, the requirements for determining the neutrino mass hierarchy with reactor neutrinos are very 
 stringent.
 
I wish to thank the organizers of NOW 2008, Prof. Fogli and Prof. Lisi, for a wonderfully stimulating atmosphere.


\end{document}